\documentclass[a4paper,11pt]{article}
\pdfoutput=1 

\usepackage{jcappub}
\usepackage[T1]{fontenc}
\usepackage{amssymb,amsmath,amsfonts,amsbsy,graphicx,microtype,rotating,bm,todonotes}
\usepackage{hyperref}
\usepackage{tcolorbox}
\usepackage{framed}

\def\be{\begin{equation}}
\def\ee{\end{equation}}
\def\ba{\begin{eqnarray}}
\def\ea{\end{eqnarray}}
\renewcommand{\(}{\left(}
\renewcommand{\)}{\right)}
\renewcommand{\[}{\left[}
\renewcommand{\]}{\right]}

\newcommand{\vp}{\varphi}
\newcommand{\mpl}{M_{\rm pl}}

\usepackage{color}
\newcommand{\tc}{\textcolor{blue}}

\usepackage[framemethod=default]{mdframed}
\newmdenv[skipabove=5pt,
skipbelow=5pt,
rightline=false,
leftline=false,
topline=false,
bottomline=false,
backgroundcolor=gray!10,
linecolor=gray,
innerleftmargin=5pt,
innerrightmargin=5pt,
innertopmargin=5pt,
innerbottommargin=5pt,
leftmargin=0cm,
rightmargin=0cm,
linewidth=4pt]{eBox}
\newmdenv[skipabove=5pt,
skipbelow=5pt,
rightline=false,
leftline=false,
topline=false,
bottomline=false,
backgroundcolor=gray!10,
linecolor=gray,
innerleftmargin=5pt,
innerrightmargin=5pt,
innertopmargin=-5pt,
innerbottommargin=5pt,
leftmargin=0cm,
rightmargin=0cm,
linewidth=4pt]{eBox2}

\title{\boldmath On the inflationary massive field with a curved field manifold}

\author[a, b]{Dong-Gang Wang}

\affiliation[a]{Leiden Observatory, Leiden University,
2300 RA Leiden, The Netherlands}
\affiliation[b]{Lorentz Institute for Theoretical Physics, Leiden University,\\
2333 CA Leiden, The Netherlands}

\emailAdd{wdgang@strw.leidenuniv.nl}

\abstract{
Massive fields during inflation provide an interesting opportunity to test new physics at very high energy scales. 
Meanwhile in fundamental realizations, the inflationary field space typically has a curved geometry, which may leave detectable imprints in primordial observables.
In this paper we study an extension of quasi-single field inflation where the inflaton and the massive field belong to a curved field manifold.
Because of the nontrivial field space curvature, the massive field here can get significant mass corrections of order the Hubble scale, thus the quasi-single field predictions on primordial non-Gaussianity are affected.
We derive the same result in an equivalent approach by using the background effective field theory of inflation, 
where a dimension-6 operator  is identified to play an important role
and its cutoff scale is associated with the curvature scale of the field space.
In addition, due to the slow-roll evolution of the inflaton, this type of mass correction has intrinsic time-dependence. 
Consequently, the running mass modifies the scaling behaviour in the squeezed limit of the scalar bispectrum, while the resulting running index measures the curvature of the internal field space.
Therefore the minimal setup of a massive field within curved field space during inflation  may naturally lead to new observational signatures of the field space geometry.
}

\keywords{\tc{inflation, effective field theory, primordial non-Gaussianity}}

\begin{document}

\maketitle
\flushbottom

\section{Introduction}
\label{intro}

Cosmic inflation, which provides a good description of the very early Universe \cite{Akrami:2018odb, Akrami:2019izv}, can also be seen as a physics laboratory at extremely high energy scales.
Through primordial perturbations, we can trace the imprints left by fundamental physics during inflation in astronomical observations, such as the cosmic microwave background (CMB) and large scale structure (LSS) surveys.
Moreover primordial non-Gaussianity, which encodes the field interactions during inflation, is believed to be one of the most powerful tools for testing new physics effects \cite{Bartolo:2004if, Chen:2010xka, Chluba:2015bqa, Meerburg:2019qqi}.
Therefore it is phenomenologically interesting to work out various non-Gaussian templates from inflation theories for future observations.
From this point of view, one well-studied example is {\it quasi-single field inflation} (QSFI) \cite{Chen:2009we, Chen:2009zp, Baumann:2011nk, Assassi:2012zq, Sefusatti:2012ye, Noumi:2012vr}, where the extra fields during inflation with mass of $\mathcal{O}(H)$ can leave unique signatures in the primordial bispectrum of curvature perturbation.
This idea has been further developed into model-independent frameworks for probing primordial physics, such as the proposals of cosmological collider physics \cite{Arkani-Hamed:2015bza, Lee:2016vti, Meerburg:2016zdz, Chen:2016uwp, Chen:2016hrz, Arkani-Hamed:2018kmz, Hook:2019zxa, Kumar:2019ebj, Wang:2019gbi} and primordial standard clocks \cite{Chen:2011zf, Chen:2014cwa, Chen:2015lza, Chen:2018cgg}.

Meanwhile, another important question in the primordial cosmology is the physical realizations of inflation in more fundamental theories.
From this theoretical perspective, one common observation is that the resulting low-energy effective theories of inflation are typically associated with a curved field manifold.
For instance, it could be the moduli field space arising in string compactifications, or the coset field space of (pseudo-)Goldstones after spontaneous symmetry breaking.
This theoretical consideration leads to the studies on {\it inflation with curved field space}. Recently there has been a revival of interest in this direction, and the representative works include $\alpha$-attractors \cite{Kallosh:2013hoa, Kallosh:2013tua, Kallosh:2013yoa, Kallosh:2015zsa} and their multi-field extensions \cite{Achucarro:2017ing, Krajewski:2018moi, Christodoulidis:2018qdw, Linde:2018hmx, Dias:2018pgj, Iarygina:2018kee}, geometrical destabilization \cite{Renaux-Petel:2015mga, Renaux-Petel:2017dia, Garcia-Saenz:2018ifx, Garcia-Saenz:2018vqf, Grocholski:2019mot, Cicoli:2019ulk}, ultra-light isocurvature scenario \cite{Achucarro:2016fby, Achucarro:2019lgo} and orbital inflation \cite{Achucarro:2019pux, Achucarro:2019mea, Welling:2019bib}, hyperinflation \cite{Brown:2017osf, Mizuno:2017idt, Bjorkmo:2019aev, Fumagalli:2019noh, Bjorkmo:2019qno}, the two-field regime of axion monodromy \cite{Pedro:2019klo}, and the analysis of new multi-field attractors \cite{Bjorkmo:2019fls, Christodoulidis:2019mkj, Christodoulidis:2019jsx}.
Usually in the curved field manifold the inflaton trajectory may demonstrate turning dynamics (or equivalently non-geodesic motion in the field space).
It has also been suggested that such kind of multi-field behaviour may be free from some possible problems faced by single field inflation \cite{Achucarro:2018vey, Bravo:2019xdo, Chakraborty:2019dfh}.
Moreover, richer phenomenology emerges in this class of multi-field models, which could be interesting for future observational detections.

Having various models of inflation with curved field space, now one may ask
 a more generic question: considering that the field space
curvature is associated to a new energy scale during inflation, then what are the {\it observational signatures of this curvature scale}?
One attempt in this direction was lately performed in Ref. \cite{Garcia-Saenz:2019njm}, where the generic cubic action is derived for the multi-field system with curved field space, and after a heavy field is integrated out, the geometrical effects manifest in the effective cubic action of curvature perturbations.
The question remains, if there are other observable imprints uniquely left by the geometry of the internal space during inflation.
\medskip

In this paper, we attack the above question in the context of QSFI, by focusing on the behaviour of a massive field living in a curved field manifold of inflation.
The main results are summarized as follows:

\begin{itemize}
\item We extend the QSFI model to the case where the inflaton and the massive field span a curved field space with a nontrivial metric.
Using the covariant formalism of multi-field inflation, we perform the background and perturbation analysis of this two-field system.
Due to the presence of the non-trivial field space curvature,
the massive field gets mass corrections which can be comparable to (or even larger than) the ``bare'' mass.
Then we provide one simple realization of QSFI with significant curved field space effects.
Furthermore,
through this concrete case study, we explicitly demonstrate how the curvature of the field space is related to an energy scale during inflation.

\item Next, we investigate the background effective field theory (EFT) of inflation with the dimension-five (dim-5) and dimension-six (dim-6) mixing operators.
This EFT approach, which has been widely adopted in the studies of massive fields, provides an alternative description for the curved field space system.
We explicitly bridge the gap between these two languages. 
In particular, the dim-6 operator in the EFT can give a significant contribution to the mass of the extra field, thus has the same effects as the curved field space. Moreover, this correction to the ``bare'' mass is essentially time-dependent, and we further consider the running behaviour of the final isocurvature mass.

\item Finally, we study the phenomenological consequences of the curved field space on QSFI predictions. As is known, the mass of the additional field leaves a unique scaling signature in the squeezed limit of the scalar bispectrum.
Here the curved field space may result in two modifications: {\it i)} the field space curvature contribution corrects the original ``bare'' mass, thus changes the predictions in the scaling index;
{\it ii)} the time-dependence of this mass correction leads to the running of the scaling in the squeezed bispectrum.
Therefore through the phenomenology of the running isocurvature mass, we can find the observational signatures left by the field space curvature.

The time dependence of the isocurvature mass $\mu$ is divided into three different regimes: running within $\mu<3H/2$ and $\mu>3H/2$, and also running through $\mu=3H/2$. To search for new predictions, we work out the modified scaling behaviour of the squeezed bispectrum caused by them one by one.
In the first two cases, the modification corresponds to the running of the scaling index in the power-law and oscillatory signals respectively, while the third case demonstrates a transition behaviour between these two types of signals.
Implications for non-Gaussianity observations are discussed.

\end{itemize}

Some of the results, for instance the field space curvature contribution to the  mass of the extra field, have been noticed in  different setups, such as geometrical destabilization \cite{Renaux-Petel:2015mga} (for negative correction, also see the early discussion in Ref. \cite{Gong:2011uw}) and spontaneous symmetry probing \cite{Nicolis:2011pv} (for positive correction).
Here we look into more generic cases of this contribution, and find it illuminating to further interpret the curved field space effects from the perspective of inflationary massive fields.
In addition, the correspondences among several different research topics are clarified.
Other results, such as the curved field space modifications to QSFI and the running phenomenology of $\mu^2$, were not discussed in the previous studies.

The outline of the paper is as follows. In Section \ref{sec:multi} we study the  massive field within a curved field space during inflation via the multi-field analysis, and demonstrate the effects of the field space curvature in a concrete example. In Section \ref{sec:eft} we take the background EFT approach to reexamine QSFI, and identify the role of a dim-6 operator and its connection with curved field space.
Section \ref{sec:pheno} focuses on the phenomenology, where the consequences of the running isocurvature mass are investigated in detail.
We summarize in Section \ref{sec:concl} with discussions on future works .

\section{When quasi-single field inflation meets a curved field space}
\label{sec:multi}

QSFI corresponds to one particular regime of inflation models, where the extra fields besides the inflaton are massive and thus generate isocurvature pertubations with a mass around the Hubble scale $H$.
The original model of QSFI in Ref. \cite{Chen:2009we, Chen:2009zp}
is described by the following matter Lagrangian
\be \label{chenwang}
\mathcal{L}_m = -\frac{1}{2}  \rho^2 (\partial \theta)^2
-\frac{1}{2} (\partial\rho)^2 - V(\rho)-V_{sr}(\theta)~,
\ee
where  the radial field $\rho$ is taken to be massive and stabilized around $\rho=\rho_0$, with $V''(\rho_0) \sim \mathcal{O}(H^2)$. Meanwhile the angular field $\theta$ plays the role of the inflaton, which is slowly rolling on a nearly flat potential along the angular direction.
In this section, we shall extend this model, and consider the situation while the inflaton and the massive field are living in a curved field manifold.

The curved field space generically arises in the low-energy effective theory of inflation, whose action with a scalar sector and Einstein gravity can be formulated as
\be \label{action}
S= \int d^4 x \sqrt{-g} \[ \frac{\mpl^2}{2} \mathbf{R} -\frac{1}{2} G_{ab}(\phi) g^{\mu\nu} \partial_\mu \phi^a \partial_\nu \phi^b -\mathcal{V}(\phi) \]~.
\ee
Notice that besides the spacetime metric $g_{\mu\nu}$, an internal field space metric $G_{ab}(\phi)$ of a non-linear sigma model also appears.
Generally speaking, the inflaton field here corresponds to one particular trajectory in the multi-dimensional field space.
Thus in addition to the adiabatic perturbations along this inflaton trajectory, the isocurvature perturbations in the orthogonal direction are also present.
To be specific, we consider an axion-dilaton system spanned by $\phi^a=(\theta, \rho)$ with the  field space metric
\be
G_{ab}=
\begin{pmatrix}
f(\rho) ~ & ~ 0 \\
0 ~ & ~ 1
\end{pmatrix}~,
\ee
which yields a non-trivial kinetic mixing for the two scalar fields.
Here the axion $\theta$ can be seen as an ``angular'' field, while the dilaton field $\rho$ corresponds to the ``radial'' direction in this internal space.
The non-trivial geometry of this internal manifold is characterized by the Ricci curvature scalar
\be \label{geomtry}
\mathbb{R}=\frac{f'(\rho)^2}{2f(\rho)^2}-\frac{f''(\rho)}{f(\rho)}~,
\ee
which is of mass dimension  $-2$.
With the choice of the potential, QSFI can be easily realized in this multi-field system\footnote{One can construct exact models of QSFI with curved field space by using the extended Hamilton-Jacobi formalism, as done in orbital inflation \cite{Achucarro:2019mea, Welling:2019bib}. This approach is not adopted here.}.
One direct extension of the original model yields the following two-field Lagrangian
\be \label{lagrangian}
\mathcal{L}_m = -\frac{1}{2}  f(\rho) (\partial \theta)^2
-\frac{1}{2} (\partial\rho)^2 - V(\rho)-V_{sr}(\theta)~,
\ee
where again $\theta$ is the inflaton and $\rho$ is the massive field.
Thus the original model can be seen as a special case of the above setup with $f(\rho) = \rho^2$,
where the field space is flat and  described by the polar coordinate.
Next, with the help of multi-field techniques, we shall investigate the QSFI with a general metric function $f(\rho)$.

\subsection{The multi-field analysis of the massive field}

For inflaton trajectories in a curved field space, the covariant formalism of multi-field inflation \cite{Gordon:2000hv, GrootNibbelink:2000vx, GrootNibbelink:2001qt, Achucarro:2010da, Gong:2011uw} provides a powerful tool to describe the background dynamics and perturbations.
Consider a turning trajectory with $\rho=\rho_0$, then the field velocity of the canonically normalized inflation is given by
$\dot\phi^2=G_{ab}\dot\phi^a\dot\phi^b=f(\rho)\dot\theta^2$
, where the dot denotes the derivative with respect to the cosmic time.
Thus we can build the tangent and normal unit vectors of this trajectory
\be
T^a \equiv \frac{\dot\phi^a}{\dot\phi}=\frac{1}{\sqrt{f(\rho_0)}}(1, 0)~,~~~~N^a=(0, 1)~.
\ee
Also the turning rate is defined as
\be \label{turning}
\Omega \equiv -N_a D_t T^a=\frac{f'(\rho_0)}{2\sqrt{f(\rho_0)}}\dot\theta~,
\ee
where $D_t$ is the covariant derivative of the field space with respect to cosmic time.
In general a geodesic trajectory in the field space yields $\Omega =0$, thus the turning parameter measures the deviation from a geodesic \cite{Achucarro:2010da}.
For the flat field metric $f(\rho)=\rho^2$, it simply yields $\Omega=\dot\theta$.
With these notations, the background equations of motion (EoMs) $D_t\dot\phi^a + 3H\dot\phi^a+V^a =0$ become
\be \label{EoMb}
\ddot\phi + 3H \dot\phi + V_T =0~,~~~~\Omega \dot\phi = V_N~,
\ee
where $V_T=T^a\nabla_a V_{sr}$ and $V_N=N^a \nabla_a V$, with $\nabla_a$ being the covariant derivative of the field space. The first equation captures the slow-roll dynamics, while the second one describes the balance between the turning and the centrifugal force.
Meanwhile the slow-roll parameters here are given by
\be \label{slowroll}
\epsilon\equiv-\frac{\dot H}{H^2}=\frac{\dot\phi^2}{2\mpl^2H^2}=\frac{f(\rho_0)\dot\theta^2}{2\mpl^2H^2}~, ~~~~\eta \equiv\frac{\dot\epsilon}{H\epsilon}=2\epsilon~.
\ee

Now let us describe the behaviour of perturbations using the background parameters above.
At the linear level, we can define the curvature perturbation $\zeta$ and the isocurvature modes $\sigma$ as
$\delta\phi^a = \sqrt{2\epsilon} \zeta T^a + \sigma N^a$.
Expanding \eqref{action} to the second order, we get the general form of the quadratic action
\be \label{quadratic}
S_2=\int d^4x  a^3\[ \epsilon\(\dot\zeta -\frac{2\Omega}{\sqrt{2\epsilon}}\sigma\)^2-\frac{\epsilon}{a^2}(\partial_i\zeta)^2+\frac{1}{2}\(\dot\sigma^2-\frac{1}{a^2}(\partial_i\sigma)^2\)-\frac{1}{2}\mu^2\sigma^2\]
~.
\ee
Here notice that at the quadratic level, the interaction between $\zeta$ and $\sigma$ is given by the turning parameter $\Omega$.
From the EoM of perturbations, this coupling corresponds to the conversion from isocurvature to curvature modes on superhorizon scales.
For a geodesic trajectory ($\Omega=0$), the curvature and isocurvature perturbations are decoupled.
In this work we mainly focus on the weakly coupled regime, {\it i.e.} $\Omega/H\ll1$.
Another interesting result of the covariant formalism is the isocurvature mass, which in general can be expressed as
\be \label{mu2}
\mu^2 = V_{NN}+\frac{1}{2}\dot\phi^2\mathbb{R}+3\Omega^2~.
\ee
Here the first term is the Hessian of the potential in the normal direction $V_{NN} = N^aN^b \nabla_a\nabla_b V$.
For the turning trajectory along the $\theta$ direction, we simply get $V_{NN}= V''(\rho_0)$, which can be seen as
the ``bare'' mass of the radial field, and is the one
usually considered in QSFI. The second and third terms are the contributions from field space curvature and turning rate.
Since we work in the weakly coupled regime with $\Omega\ll H$, thus the last term contribution can be neglected for $\mu^2\sim \mathcal{O}(H^2)$.

The main focus of this paper is the second term in \eqref{mu2}. This field space curvature contribution  can be tracked back to the kinetic term of the the two-field system in \eqref{lagrangian}.
Naively speaking, when we derive the perturbed Lagrangian,
the $\sigma$ field mass has contributions from the second order expansion of $f(\rho)$, which is related to the Ricci scalar in \eqref{geomtry}.
This is a unique correction in the quantum field theory with time-dependent background. Thus for inflation, it is always accompanied by the inflaton field velocity $\dot\phi^2$, whose magnitude can be estimated from the current observations\footnote{In the weakly coupled regime of QSFI, since the massive field correction to the final power spectrum of $\zeta$ is small, the single field prediction $P_{\zeta}=H^4/(4\pi^2\dot\phi^2)$  remains valid approximately. Then from the observational result $P_\zeta\simeq 2 \times 10^{-9}$ \cite{Akrami:2018odb}, one gets $\dot\phi^2\simeq 10^7 H^4$.}: $\dot\phi^2\simeq 10^7 H^4$.
Therefore unless the field space curvature is extremely small, the second term in \eqref{mu2} should not be neglected.

In the following we shall demonstrate in a case study that the field space curvature is typically associated with a new energy scale during inflation, and for natural choices of this scale, the curvature term in $\mu^2$ can be comparable to or even larger than the $V''$ term.

\subsection{A concrete example: inflation in coset space}
\label{sec:coset}

In order to avoid the $\eta$ problem \cite{Copeland:1994vg}, the low-energy effective theories of inflation are usually supposed to be described by (pseudo-)Goldstone bosons protected by an (approximate) internal symmetry, such that the slow-roll potential is free from quantum corrections.
Consequently the inflaton may roll in a {\it non-abelian coset space} $G/H$ defined by the symmetry breaking pattern.
While the details of a relevant project will be presented in a future paper \cite{Wang:2020aaa},  here let us look at two simplest cases of coset space with nontrivial geometries which have been considered before in Ref.~\cite{Burgess:2014tja, Klein:2017hch}.

\begin{figure}[tbhp]
\centering
\includegraphics[width=0.7\linewidth]{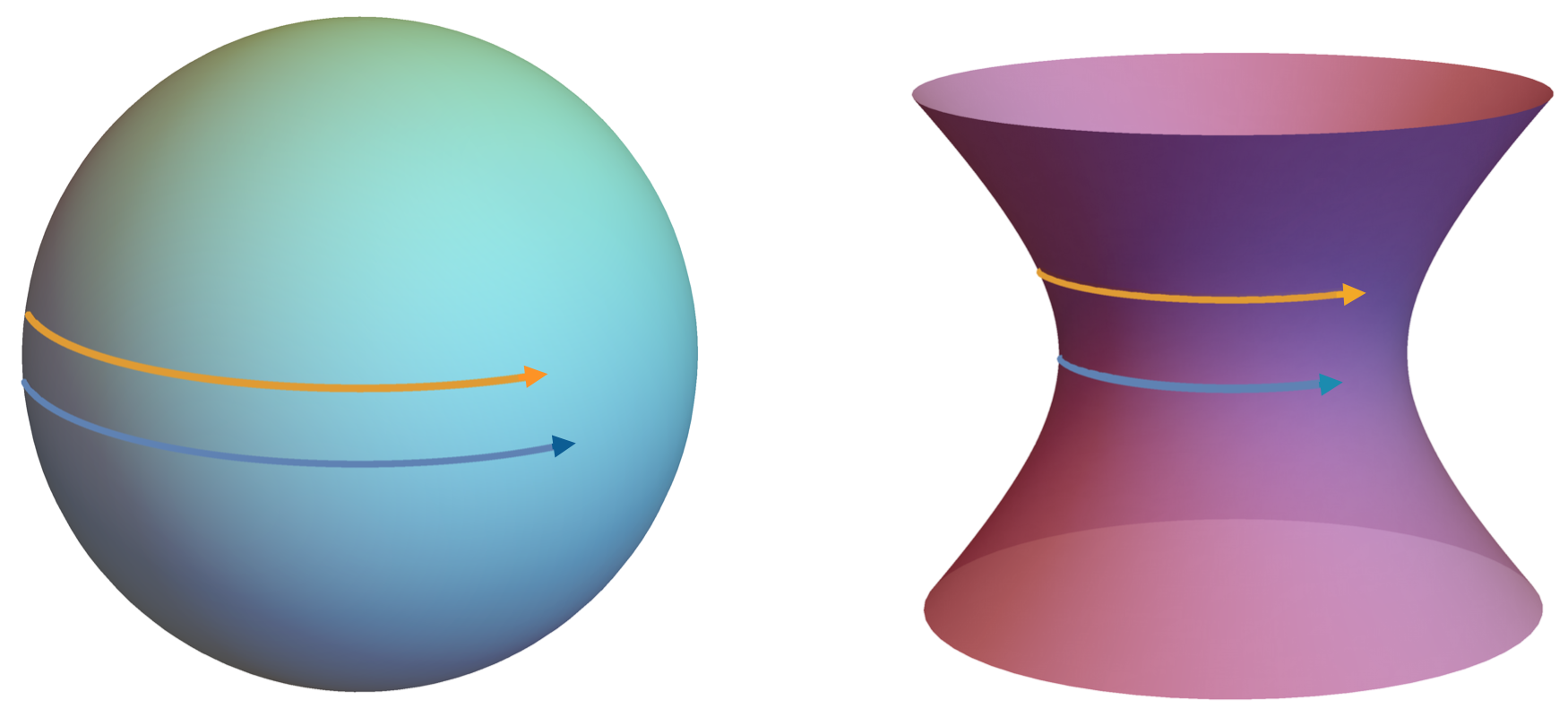}
\caption{The $SO(3)/SO(2)$ (left) and $SO(2,1)/SO(2)$ (right) coset spaces, with the corresponding geodesic trajectories (blue curves) and examples of possible deviations (orange curves).}
\label{fig:coset}
\end{figure}

\begin{itemize}
\item $SO(3)/SO(2)$. This coset space is a 2d-sphere  defined by $\phi_1^2+\phi_2^2+\phi_3^2=R^2$ in the three-dimensional Euclidean space, where the constant $R$ is the radius of the spherical surface. Thus
it is convenient to use the spherical coordinates
\be
\phi_1=R\cos\varrho\cos\theta~,~~~~\phi_2=R\cos\varrho\sin\theta~,
~~~~\phi_3=R\sin\varrho~,
\ee
where $\theta$ and $\varrho$ are two Goldstone fields in the coset.
As a result, the line element of this field space becomes
$ds^2 = R^2 \left(d\varrho^2+ \cos^2\varrho d\theta^2\right)$. If we canonically normalize $\varrho$ by redefining $\rho = R \varrho$, then the kinetic term of the two Goldstones is expressed as
\be \label{spherical}
K= -\frac{1}{2} (\partial\rho)^2 -\frac{1}{2}  R^2 \cos^2\(\frac{\rho}{R}\) (\partial \theta)^2~,
\ee
which corresponds to the system in \eqref{lagrangian} with $f(\rho)= R^2 \cos^2({\rho}/{R})$.
This field space has a positive constant curvature with $\mathbb{R}=2/R^2$.

\item $SO(2,1)/SO(2)$.
This non-compact coset yields a 2d-hyperbola defined by $\phi_1^2+\phi_2^2-\phi_3^2=R^2$ in the three-dimensional Minkowski space, as shown in Fig.~\ref{fig:coset}.
We use the following field coordinates
\be
\phi_1=R\cosh\varrho\cos\theta~,~~~~\phi_2=R\cosh\varrho\sin\theta~,
~~~~\phi_3=R\sinh\varrho~.
\ee
Again the coset space is spanned by the Goldstone fields $\theta$ and $\varrho$, with the line element $ds^2 = R^2 \left(d\varrho^2+ \cosh^2\varrho d\theta^2\right)$. Using the field redefinition $\rho = R \varrho$, we get the Goldstone kinetic term as
\be \label{hyperbola}
K= -\frac{1}{2} (\partial\rho)^2 -\frac{1}{2} R^2 \cosh^2\(\frac{\rho}{R}\) (\partial \theta)^2~,
\ee
which has $f(\rho)= R^2 \cosh^2({\rho}/{R})$. This is a hyperbolic space\footnote{One can connect this with the hyperbolic field space in $\alpha$-attractors with $\mathbb{R}=-2/(3\alpha)$, and there $\alpha$ is related to the radius of curvature by $\alpha= R^2/3$.} with a negative constant curvature given by $\mathbb{R}=-2/R^2$.

\end{itemize}

As we see from these two examples, the Ricci scalar is determined by the radius $R$ of the field space, which corresponds to the symmetry breaking scale in this setup.
Now we take into account the motion of the inflaton by assuming a slow-roll potential which softly breaks the shift symmetry.
Let us first consider the geodesic trajectories in these field spaces, which can be related to the spontaneous symmetry probing solutions discussed in Ref.~\cite{Nicolis:2011pv}.

In the $SO(3)/SO(2)$ coset, the geodesic is a trajectory along the maximal circle, as shown by the blue curve in the left panel of Fig.~\ref{fig:coset}. Here without losing generality we take it to be the equator with $\rho =0$, and the canonically normalized inflaton $\phi$ is driven by a slow-roll potential in the $\theta$ direction, with $\phi=R\theta$.
If there is no explicit symmetry breaking for the $\rho$ field, naively this Goldstone is supposed to be massless. However, because of the rolling of another Goldstone $\theta$, the ``not-rolling'' Goldstone $\rho$ acquires  a mass \cite{Nicolis:2011pv}
\be
m_\rho^2 = \dot\theta^2 = \frac{\dot\phi^2}{R^2}~,
\ee
which is exactly the second term in \eqref{mu2}. From here we can explicitly see that, the curved field space contribution to the isocurvature mass is associated with the internal curvature scale. For $R \sim \dot\phi /H \simeq 3600 H$,  this contribution is $\mathcal{O}(H^2)$; while for $R\sim\mpl$ it becomes slow-roll suppressed as $\sim \epsilon H^2$.

For the non-compact coset  $SO(2,1)/SO(2)$, let us consider  the inflaton trajectory that is also along the $\theta$ direction, then for the field space in \eqref{hyperbola}
the geodesic motion is given by $\rho=0$, which is the blue curve in the right panel of Fig.~\ref{fig:coset}. Here the $\rho$ field also acquires a similar mass correction from the rolling of the inflaton in the hyperbolic field space. But this is a tachyonic contribution $-\dot\phi^2/R^2$, since the field space curvature is negative.
Thus to stabilize the isocurvature perturbation during inflation, one needs to break the shift symmetry and engineer a potential in the $\rho$ direction.

Now we consider small deviations from the geodesics, for which $\Omega/H\ll 1$ and thus the curvature and isocurvature perturbations are weakly coupled. This can be easily achieved by perturbing the above geodesics away from the equator\footnote{A potential in the $\rho$ direction is needed for this type of toy model trajectories. Here we keep agnostic about the specific form of the potential, and consider the consequences of this non-geodesics motion directly.}, such as the orange trajectories in Fig.~\ref{fig:coset}.
For the spherical space case, the trajectory is taken to be the latitude line $\rho =  \delta $, where $\delta$ parametrizes the deviation.
Then \eqref{turning} yields $\Omega\simeq -(\delta/R) \dot\theta$ which can be much smaller than $H$ for $\delta\ll R$.
Similarly in the hyperbolic field space, a non-geodesic trajectory with $\rho = \delta$ yields $\Omega\simeq (\delta/R) \dot\theta$.
Since these deviations from the geodesics are kept to be small, the field space curvature contribution to $\mu^2$ discussed above remains valid.
Therefore these isometry trajectories in the coset space provide simple realizations of QSFI with curved field manifold.

\

In summary, from the above example we identify that $\mathbb{R}\sim 1/R^2$, where the curvature radius $R$ can be seen as the energy scale describing the curved field space geometry. Moreover, it may lead to significant contribution to the isocurvature mass
\begin{eBox}
\be \label{isomass1}
\mu^2
\simeq V''(\rho_0) + \frac{\dot\phi^2}{2}\mathbb{R} = V''(\rho_0) \pm  \frac{\dot\phi^2}{R^2}~,
\ee
\end{eBox}
which should not be neglected. For certain ranges of the curvature scale, this correction could be comparable to or even larger than the Hubble scale, which may dominate $\mu^2$ in QSFI.
As a result, the model predictions of QSFI for primordial non-Gaussianities would be affected, which we shall explore in detail in Section \ref{sec:pheno}.

Before concluding this section, we would like to mention another interesting observation: in \eqref{mu2} the second term $\dot\phi^2\mathbb{R}/2$ is {\it time-dependent} during inflation, since the inflaton field velocity $\dot\phi^2= 2 \epsilon H^2 \mpl^2$ is evolving.
Although it is a small effect, when the field space curvature contribution is non-negligible, we may expect running behaviour for the isocurvature mass, which we shall describe in detail at the end of the next section.

\section{The EFT of background fields revisited}
\label{sec:eft}

In this section we reexamine
QSFI via the background EFT of inflation.
Usually to achieve the slow-roll evolution and the nearly scale-invariant  primordial perturbations, the inflaton field is believed to be protected by an (approximate) shift-symmetry.
Based on this argument, one can construct {the EFT of background fields} for inflation without the knowledge of microphysical realizations \cite{Weinberg:2008hq}, which provides a  model-independent framework for studying physics in the primordial Universe.

Here we are mainly interested in the coupling between the extra-fields and the inflaton (denoted as $\varphi$ in this section).
Since the massive field $\rho$ does not respect the shift symmetry, the leading contribution to the mixing between the inflaton and $\rho$ is given by a dimension-five (dim-5)  operator in the EFT expansion
\be \label{dim5}
\mathcal{L}_{\rm int}^5 = -\frac{1}{2\Lambda_1} (\partial \varphi)^2 \rho~,
\ee
where $\Lambda_1$ is the cutoff scale.
This operator, which has been elaborately investigated in the studies of QSFI and related topics \cite{Assassi:2013gxa, Arkani-Hamed:2015bza, An:2017hlx, Tong:2017iat}, is the leading order term in the EFT expansion.
Realistically higher order terms should also be present.
In the following we shall show how one could connect {\it the background EFT} with the {\it curved field space} in QSFI, and then focus on the role of a dimension-six (dim-6) operator
\be \label{dim6}
\mathcal{L}_{\rm int}^6 = \pm\frac{1}{2\Lambda_2^2} (\partial \varphi)^2 \rho^2~,
\ee
which can introduce the same effects as the field space curvature.
The connection has also been noticed in geometrical destabilization \cite{Renaux-Petel:2015mga}, while in the current work we bridge the gap explicitly and highlight the generic effects for massive fields.

\subsection{Bridging the background EFT with the curved field space}

Let us begin with the following EFT Lagrangian of two background fields
\be \label{EFT}
\mathcal{L}_m = -\frac{1}{2}\(1 + c_1 \frac{\rho}{\Lambda} + c_2 \frac{\rho^2}{\Lambda^2} \)(\partial\varphi)^2 - \frac{1}{2} (\partial\rho)^2 - \frac{1}{2}m^2 \rho^2 -V_{\rm sr}(\varphi)~.
\ee
where $\Lambda$ is an overall cutoff for the dim-5 and dim-6 operators. The dimensionless coefficients $c_1$ and $c_2$ with $|c_{1,2}|\leq1$ are introduced to represent their relative size and signs, thus $\Lambda_1 = \Lambda/c_1$ and $\Lambda_2 = \Lambda/\sqrt{|c_2|}$.
Furthermore these two mixing operators are considered to be perturbative corrections to the single field slow-roll inflation, {\it  i.e.} $\rho/\Lambda \ll 1$.
Notice that the system has the same form with \eqref{lagrangian}, while
in the curved field space language these two operators yield a non-trivial field space metric function
\be
f(\rho) = 1 + c_1 \frac{\rho}{\Lambda} + c_2 \frac{\rho^2}{\Lambda^2}~.
\ee
Thus the EFT in \eqref{EFT} can be seen as the  expansion of a curved manifold Lagrangian around a fixed trajectory with constant $\rho$.
From \eqref{geomtry} we also get the Ricci curvature as
\be
\mathbb{R}
\simeq  -\frac{2c_2-c_1^2/2}{\Lambda^2} + \mathcal{O}\( \frac{\rho}{\Lambda}\)~.
\ee
As we see, since the curvature contains the second order derivative of the metric, the dim-6 operator will play a role here in general.

First let us look at the background dynamics.
The EoM of the $\rho$ field, which is the centrifugal force equation in \eqref{EoMb}, yields the stabilized value
for the massive field at $\rho = \rho_0$\footnote{Note here $\rho_0$ depends on the inflaton velocity, thus strictly speaking it is not a constant. See Ref.~\cite{Achucarro:2019mea, Welling:2019bib} for models with exactly constant $\rho_0$.}
\be \label{rho0}
\frac{1}{2}\(c_1\frac{1}{\Lambda} +2 c_2 \frac{\rho_0}{\Lambda^2}\) \dot\varphi^2 = m^2\rho_0 ~~~~ \Rightarrow ~~~~
\frac{\rho_0}{\Lambda} =\frac{c_1}{2} \frac{\dot\varphi^2/\Lambda^2}{m^2-c_2\( \dot\varphi^2/\Lambda^2\)}~.
\ee
To ensure the validity of EFT, one needs $\rho_0/\Lambda \ll 1$.
Then the canonically normalized inflaton is just $\phi=f(\rho_0)\varphi$ with $f(\rho_0)\simeq1$.
Also there is a turning rate given by\footnote{In the expansion we also consider the possibility for a hierarchy between $c_1$ and $c_2$, such as $c_1\sim c_2 (\rho_0/\Lambda) $.}
\be
\Omega^2
\simeq \frac{1}{4} \(c_1 + 2 c_2 \frac{\rho_0}{\Lambda} \)^2 \frac{\dot\varphi^2 }{\Lambda^2} + \mathcal{O}\( \frac{\rho_0}{\Lambda}\)~,
\ee
thus the curvature and isocurvature perturbations are coupled at the linear level.
The weak coupling condition $\Omega/H\ll1$ here implies that $|c_1+ 2 c_2 ({\rho_0}/{\Lambda})|(\dot\varphi/\Lambda) \ll 2H$.
The isocurvature mass follows from \eqref{mu2} as
\be
\mu^2 \simeq m^2 +\[c_1^2 -c_2 + 3 c_2^2 \(\frac{\rho_0}{\Lambda} \)^2 +3c_1 c_2\(\frac{\rho_0}{\Lambda} \) \]\(\frac{\dot\varphi}{\Lambda}\)^2
\simeq  m^2 +\(\frac{c_1^2}{4} -c_2 \)\(\frac{\dot\varphi}{\Lambda}\)^2 ,
\ee
where in the second approximation the weak coupling condition is used. Here the $\rho$ field ``bare'' mass $m^2$  gets corrections from the mixing operators due to the time-dependent background of the inflaton field.

\begin{figure}[tbhp]
\centering
\includegraphics[width=0.6\linewidth]{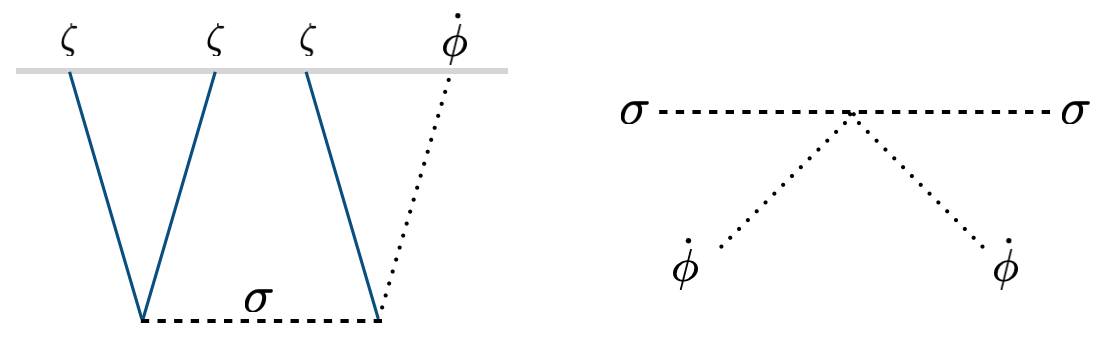}
\caption{Different roles of dim-5 and dim-6 operators shown in Feynman diagrams.}
\label{fig:feynmann}
\end{figure}

Now we comment on the role of the dim-5 operator.
At the background level,
this operator contains a tadpole for $\rho$  which contributes to stabilize the massive field .
If we switch it off by setting $c_1=0$, the centrifugal force equation \eqref{rho0} yields $\rho_0 =0$ and the trajectory becomes a geodesic with $\Omega=0$.
Thus this operator is important for the non-geodesic motion of the inflaton, and  leads to the mixing between curvature and isocurvature perturbations. Also it contributes to the left diagram in Fig.~\ref{fig:feynmann} for the scalar bispectrum in QSFI, thus its size can be related to the amplitude of the non-Gaussian signals.
Meanwhile the cutoff scale of this operator is constrained in the weakly coupled regime.
If we turn off the dim-6 operator by setting $c_2=0$, the weak coupling condition  $|c_1|(\dot\varphi/\Lambda) \ll 2H$ yields a lower bound on  $\Lambda_1 = \Lambda/c_1$.
As a result, the dim-5 operator itself gives negligible corrections to the isocurvature mass which simply reduces to the ``bare'' one $\mu^2 \simeq m^2$.

\subsection{On the role of the dimension-6 operator}

As a next-to-leading order correction, usually the dim-6 operator in \eqref{dim6} is supposed to be sub-dominant.
For instance, it contributes to the scalar bispectrum through loop diagrams, which turns out to be negligible \cite{Arkani-Hamed:2015bza, Kumar:2019ebj}.
But as discussed above (and also pointed out in Ref.~\cite{Renaux-Petel:2015mga, Kumar:2019ebj, Fumagalli:2019ohr} previously), this operator may give nontrivial corrections to the isocurvature mass.
One can interpret this effect by the right diagram in Fig.~\ref{fig:feynmann}, where the two inflaton legs are taken to be the background, and then the extra field get mass corrections through this operator.

To show its effects explicitly, let us consider the situation where the massive field enjoys an approximate $\mathbb{Z}_2$ symmetry, such that the coefficients of two mixing operators satisfy
\be
|c_1|  \ll |c_2| = 1 .
\ee
Thus the overall cutoff scale $\Lambda$ coincides with the one for the dim-6 operator $\Lambda_2$, and there is a hierarchy: $\Lambda_1\gg\Lambda_2$.
The Ricci scalar becomes $\mathbb{R}\simeq  -{2c_2}/{\Lambda^2} $.
In the curved field space analogy, this case corresponds to the  small deviations from the equator ($\rho=0$) in the coset space discussed in Section \ref{sec:coset}.
For instance, in the $SO(3)/SO(2)$ system \eqref{spherical}, suppose the massive field is stabilized around $\rho=\rho_b\ll R$ by a potential $V(\rho)=m^2(\rho-\rho_b)^2/2$.
If we look at a local patch of the spherical surface around the latitude $\rho=\rho_b$, then the metric function there can be expanded as
\be
R^2\cos^2\(\frac{\rho}{R}\) = R^2\[ 1 -2\frac{\rho_b}{R} \frac{\rho-\rho_b}{R} -\frac{(\rho-\rho_b)^2}{R^2} \]  +...
\ee
By redefining $\rho-\rho_b\rightarrow\rho$ and $R\theta\rightarrow \varphi$, we  recover the EFT Lagrangian in \eqref{EFT} with $\Lambda=R$, $c_1=-2\rho_b/R$ and $c_2=-1$.
Similarly the $SO(2,1)/SO(2)$ example can also be formulated into the Background EFT language, with $\Lambda=R$ as well, but $c_1=2\rho_b/R$ and $c_2=1$ instead.
As we see here, {\it the cutoff scale $\Lambda$ of the dim-6 operator plays the role of the curvature scale of the field space}, while the sign of $c_2$ denotes if it is positively or negatively curved.

The background dynamics of this hierarchical system follows directly.
Because of the small but nonzero dim-5 operator, the massive field gets deviated from the potential minimum by the centrifugal force \eqref{rho0}.
Since $|c_1| \ll 1$, we find it easy to guarantee $\rho_0/\Lambda\ll1$, and thus the EFT description is justified.
Notice that, while the size of the dim-5 operator is still constrained by the weak coupling condition, there is more freedom for the cutoff scale of the dim-6 operator.
One particularly interesting regime for QSFI is for $\dot\varphi^2/\Lambda^2\sim H^2$, which corresponds to $\Lambda \sim 3000 H$ by considering $\dot\varphi^2\simeq\dot\phi^2\simeq 10^7 H^4$.
The isocurvature mass becomes
\begin{eBox}
\be \label{isomass}
\mu^2
\simeq m^2 -  c_2 \(\frac{\dot\varphi}{\Lambda}\)^2   = m^2 - 2 c_2 \epsilon H^2 \(\frac{\mpl}{\Lambda}\)^2~,
\ee
\end{eBox}
where in the second equality we introduced the slow-roll parameter $\epsilon$.
The second term, which represents the contribution from the dim-6 operator, can be comparable with (or even larger than) the ``bare'' mass $m^2$, and thus should not be neglected.
This expression has the same form with \eqref{isomass1} in terms of the field space curvature.
Thus for $c_2=1$, this correction is negative, which is equivalent to the hyperbolic field space case; while $c_2=-1$ makes the massive field heavier and corresponds to the positively curved field space.

Moreover, as we briefly mentioned at the end of Section \ref{sec:multi}, this correction to $\mu^2$ is time-dependent due to the rolling behaviour of the inflaton\footnote{This is similar to what happens for geometrical destabilization \cite{Renaux-Petel:2015mga}, where the running isocurvature mass with negative $\mathbb{R}$ may lead to tachyonic instability, which ends inflation prematurely or initiates a sidetracked phase \cite{Garcia-Saenz:2018ifx, Grocholski:2019mot}. Here instead of $\mu^2<0$, we are interested in both the positive- and negative-running behaviour in the stable (QSFI) regime with $\mu^2\sim\mathcal{O}(H^2)$, and mainly focus on its effects on the large-scale modes which can be probed by CMB or LSS surveys.}.
Now let us formulate the running behaviour of the isocurvature mass more specifically.
If we consider the evolution of $\epsilon$ from the time denoted by the number of e-folds $N_l$, then as long as $N-N_l$ is not too big we have
\be
\epsilon(N)  \simeq \epsilon(N_l) +  \epsilon'(N_l) (N-N_l)
= \epsilon_l \[ 1 + \eta_l(N-N_l)\] ,
\ee
with $\epsilon_l = \epsilon(N_l)$ and $\eta_l=\eta(N_l)$.
Correspondingly the time dependence of $\mu^2$ in \eqref{isomass} can be parametrized as
\begin{eBox}
\be \label{runningmass}
\mu^2 (N)
= \mu_l^2 + \lambda  (N-N_l) H^2~,
\ee
\end{eBox}
where $\mu_l^2 =  m^2 - 2 c_2 \epsilon_l H^2 ({\mpl}/{\Lambda})^2$ is the isocurvature mass at the time of $N_l$, and $\lambda$ is the running parameter expressed in both the curved field space and the EFT languages as
\begin{eBox}
\be \label{lambda}
\lambda =  \eta_l \frac{\dot\phi^2}{2H^2}\mathbb{R} = - \eta_l\frac{c_2}{H^2} \frac{\dot\vp^2}{\Lambda^2}  ~.
\ee
\end{eBox}
As expected, typically the time-dependence is small and suppressed by $\eta_l$.
We estimate the size of  $\lambda$ in the conformal limit of inflation $\epsilon\ll \eta$ \cite{Pajer:2016ieg}.
This hierarchy between slow-roll parameters\footnote{It can be achieved by taking $V_{\rm sr}(\varphi)$ to be the plateau-like potentials, such as Starobinsky inflation \cite{Starobinsky:1980te} and $\alpha$-attractors \cite{Kallosh:2013hoa, Kallosh:2013tua, Kallosh:2013yoa, Kallosh:2015zsa}, where the slow-roll parameters evolve as $
\epsilon  \sim  {1}/{N^2}$ and $\eta \simeq {2}/{N}$.} is indicated by the observational upper bound on the tensor-to-scalar ratio $r=16\epsilon < 0.064$ and the observed value of the scalar tilt $n_s=1-\eta -2\epsilon=0.9649\pm0.0042$ \cite{Akrami:2018odb}, which also yields $\eta\simeq0.035$.
Then the running parameter in \eqref{lambda} is mainly controlled by the cutoff scale $\Lambda$. For $\Lambda \sim 3000 H$, and thus $\dot\varphi^2/(H^2\Lambda^2) \sim \mathcal{O}(1)$, we get $|\lambda|\lesssim 0.1$.
In principle a larger running can be achieved by lowering $\Lambda$, but then for the interest of QSFI one may need fine tune the correction in \eqref{isomass} against the ``bare'' mass such that $\mu^2\sim \mathcal{O}(H^2)$.

\section{Phenomenology of the running isocurvature mass}
\label{sec:pheno}

With the above analysis of massive fields living in curved field space,
the goal of this section is to investigate the phenomenological consequences of a nontrivial field space curvature in QSFI, and focus on the effects of the running isocurvature mass.

First of all, let us briefly review the phenomenology of QSFI.
To characterize primordial non-Gaussianity, the bispectrum of curvature perturbation is usually defined as
\be \label{bispectr}
\langle \zeta_{\bf k_1} \zeta_{\bf k_2} \zeta_{\bf k_3} \rangle
\equiv (2\pi)^3 \delta^{(3)}({\bf k_1+k_2+k_3}) B_\zeta(k_1, k_2, k_3)~.
\ee
We are particularly interested in the squeezed configurations of the momentum triangles formed by two short modes $k_1=k_2=k_s$ and one long mode $k_3=k_l$, with $k_l\ll k_s$.
One  of the most interesting results in QSFI is that,
the scaling behaviour of the bispectrum in this squeezed limit is uniquely determined by the isocurvature mass as follows \cite{Chen:2009zp, Baumann:2011nk, Noumi:2012vr}
\ba
&&\lim_{k_l\ll k_s} B_\zeta \propto \frac{1}{k_l^3k_s^3} \(\frac{k_l}{k_s}\)^{3/2-\nu}
~~~~~~~~~~~~~~~~~~~~~~~~~~~~{\rm for} ~~
\mu < 3H/2~, \label{underscaling} \\
&&\lim_{k_l\ll k_s} B_\zeta \propto \frac{1}{k_l^3k_s^3} \(\frac{k_l}{k_s}\)^{3/2}
\cos\[ i{\nu}\ln\(\frac{k_l}{k_s}\) + \delta_\nu \]
~~~~{\rm for} ~~
\mu > 3H/2~. \label{overscaling}
\ea
where the scaling index $\nu$ is a function of the isocurvature mass
\be \label{nu}
\nu = \sqrt{\frac{9}{4}-\frac{\mu^2}{H^2}}~,
\ee
and $\delta_\nu$ is a phase factor depending on $\nu$.
Here $\mu=3H/2$ is the critical mass which divides the isocurvature mass spectrum into light and heavy regimes.
Notice that in our notation, $0<\nu<3/2$ for $\mu < 3H/2$, and it becomes imaginary when $\mu>3H/2$.
Therefore through this observational channel of non-Gaussianities, one can measure the mass of the additional field in a model-independent manner.

Intuitively, the above scaling can be understood from the superhorizon behaviour of the massive field \cite{Baumann:2011nk}.
During inflation, the EoM of the isocurvature mode is
\be \label{EoM}
\sigma_k ''+ \frac{k^2}{a^2H^2} \sigma_k + 3 \sigma_k ' + \frac{\mu^2}{H^2} \sigma_k = 0~,
\ee
where primes denote derivatives with respect to the number of e-folds $N$.
For QSFI, we are mainly interested in the regime $\mu/H\sim \mathcal{O}(1)$.
Thus on the superhorizon scales ($k\ll aH$), the second term is sub-dominant compared with the mass term. Approximately the EoM above becomes the one for a damped oscillator $\sigma_k '' + 3 \sigma_k ' + ({\mu^2}/{H^2}) \sigma_k \simeq 0$, and for a constant $\mu^2$ it has two decaying solutions
\be
\sigma_k (N)
\propto e^{-(3/2\pm\nu)(N-N_k)}~.
\ee
Here $N_k$ is the e-folds when $\sigma_k$ mode exits the horizon.
For the light field case ($\mu < 3H/2$), the solution with minus sign dominates, and it corresponds to the underdamped decay.
For the heavy field case ($\mu > 3H/2$), the imaginary $\nu$ leads to the overdamped oscillations.
Now we consider the modulation of a long wavelength mode ($k_l$) on the short wavelength modes ($k_s$).
Suppose that the $k_l$-mode exits the horizon at $N_l$, then later when
the $k_s$-modes  exit the horizon at $N_s$, the amplitude of the $k_l$-mode already decays by
\be \label{modulat}
\sigma_{k_l} (N_s) =  \sigma_{k_l} (N_l)  e^{-(3/2\pm \nu)(N_s - N_f)}
=\sigma_{k_l} (N_l)\( \frac{k_l}{k_s}\)^{3/2\pm \nu}~,
\ee
where in the second equality $e^{N_s - N_f} = k_s/k_l$ is used.
As a result, the modulation of the long wavelength mode on the $k_s$ modes will inherit this decayed amplitude.
When $\mu < 3H/2$, the decaying solution with minus sign gives the power-law scaling in \eqref{underscaling}. While for $\mu > 3H/2$, $\nu$ is imaginary and the scaling can be written into the oscillatory form in \eqref{overscaling}.

For QSFI with curved field space, we first notice that the scaling behaviour of the squeezed bispectrum is determined by the full isocurvature mass. Thus when the field space curvature contribution to $\mu^2$ is significant,
what we measure in the non-Gaussianity observation is no longer the ``bare'' mass of the additional field,
since the scaling index $\nu$ in \eqref{nu} is changed by this nontrivial mass correction\footnote{Similar correction has been noticed in the context of cosmological colliders as a contamination \cite{Chen:2016uwp, Chen:2016hrz}.}.
It is also possible that in $\mu^2$ the dominant contribution comes from the field space curvature term. For instance, in the example of inflation in coset space discussed in Section \ref{sec:coset}, the Goldstone field in the normal direction of the inflaton trajectory has a zero ``bare'' mass, but can still become massive due to the curved field space effect.
In this sense the QSFI predictions based on the ``bare'' mass will be corrected, though it is difficult to distinguish these two mass contributions from each other.
However, the time-dependence of the field space curvature term may break the degeneracy, which we shall study in the rest of this section.

Let us take the parametrization in \eqref{runningmass} as  our starting point.
Before horizon-exit, since in \eqref{EoM} the second term dominates, the evolution of $\sigma_k$ is barely affected by the running mass.
Thus the conventional mode function with a Bunch-Davies initial condition provides a good description.
This is also shown in the shaded regions of Figs. \ref{fig:running} and \ref{fig:running2},
where the full numerical solutions with running mass
agree with dashed grey curves very well in the subhorizon regime.
But the superhorzion evolution of  $\sigma_k$ differs from the conventional case with a constant mass.
For simplicity, we define a rescaled mode function as
\be
\widetilde{\sigma}_k  = e^{3N/2} \sigma_k ~.
\ee
Then for the $k_l$ mode which exits the horizon at e-folds $N_l$, the superhorizon EoM can be approximately written into the following form
\be \label{EoMr}
\widetilde{\sigma}_{k_l} '' - \[\nu_l^2 - \lambda   (N-N_l) \] \widetilde{\sigma}_{k_l} = 0~,
\ee
where the scaling index at $N_l$ is given by the isocurvatue mass at that time $\nu_l^2= {9}/{4} - {\mu_l^2}/{H^2}$.
This equation has the following analytical solution with two Airy functions
\be \label{airy}
\widetilde\sigma_{k_l} =  C_1 {\rm Ai} \[ \frac{\nu_l^2 - \lambda  (N-N_l)}{(-\lambda )^{2/3}} \] +  C_2 {\rm Bi} \[ \frac{\nu_l^2 - \lambda  (N-N_l)}{(-\lambda )^{2/3}} \]~,
\ee
where $C_1$ and $C_2$ are two integration constants determined by the initial condition.
In the following we shall explore the behaviour of this solution in three different regimes, and their modification on the scaling of the squeezed bispectrum.

\subsection{Running in the $\mu <3H/2$ regime}

\begin{figure}[tbhp]
\centering
\includegraphics[width=0.475\linewidth]{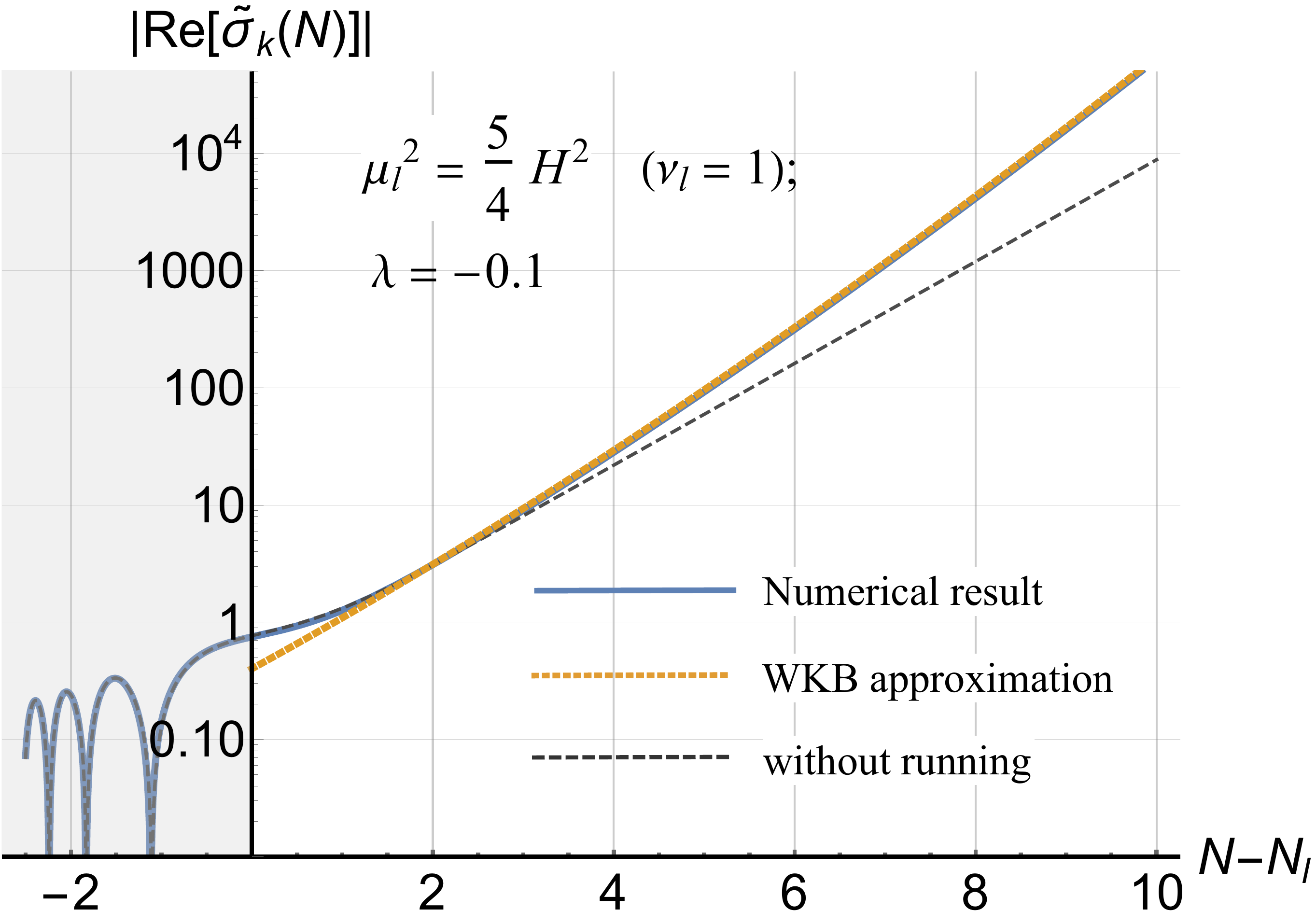}
\includegraphics[width=0.475\linewidth]{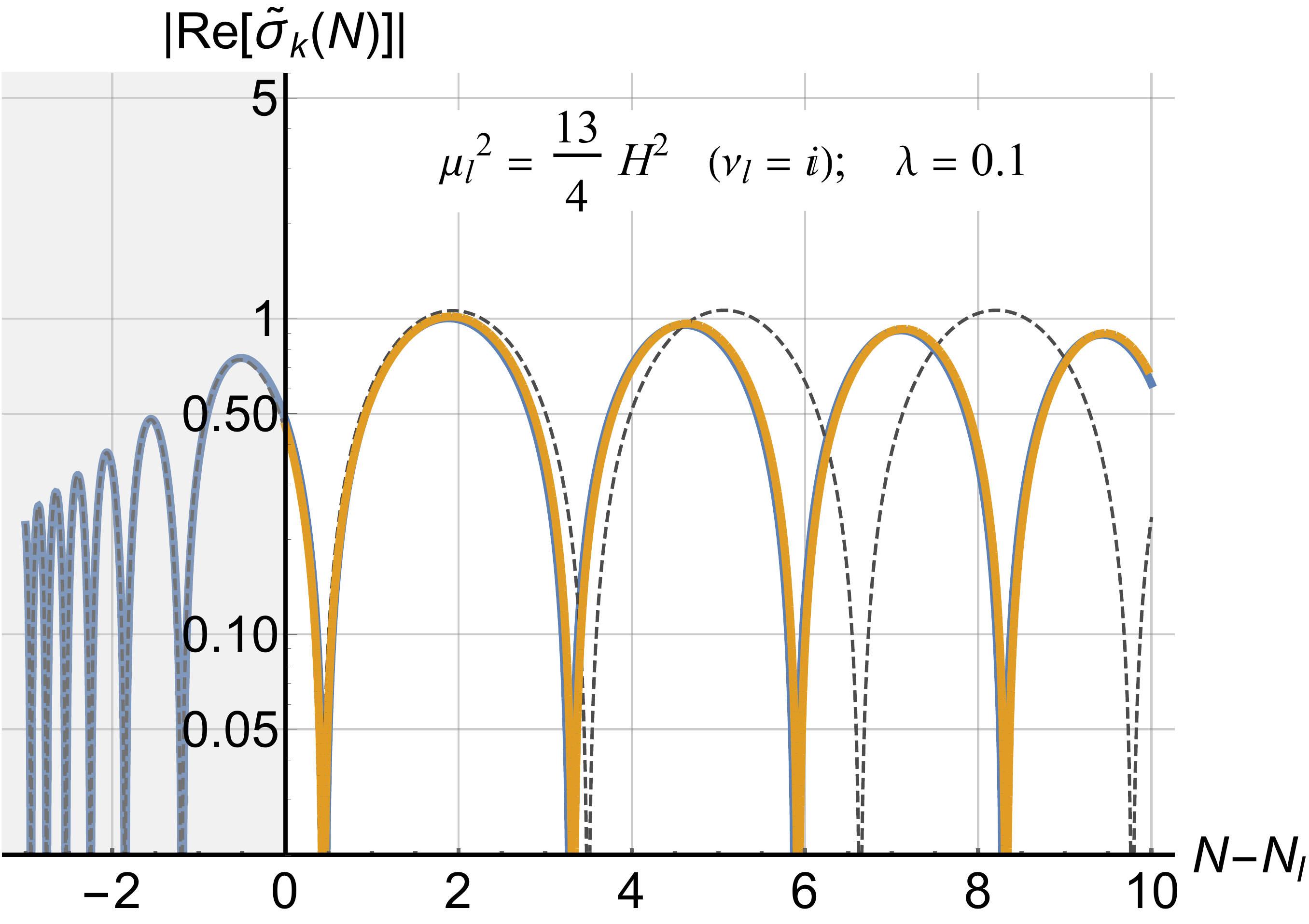}
\caption{The evolution of the rescaled isocurvature mode function for cases with: negative-running mass in the $\mu<3H/2$ regime (left panel) and positive-running mass in the $\mu>3H/2$ regime (right panel). In both figures, the blue curves are the full numerical solutions of $\tilde{\sigma}_k$, the orange dotted lines are the WKB approximation on superhorizon scales, and the dashed grey lines are the solutions with constant masses ($\lambda=0$). The shaded parts correspond to the subhorizon regime.}
\label{fig:running}
\end{figure}

First, let us look at the situation where the isocurvature mass is running in the $\mu <3H/2$ regime, which means $\nu_l^2 - \lambda  (N-N_l)>0$.
Since $\mu^2$ varies slowly, in most cases the EoM \eqref{EoMr} can be solved by the WKB approximation.
More specifically the adiabatic condition here is given by
\be
\nu_l^2 - \lambda  (N-N_l)\gg (|-\lambda |)^{2/3}~.
\ee
This breaks down when the isocurvature mass is running close to  $3H/2$, which we leave for consideration in Section \ref{sec:through}.
Then the first order WKB solution of the rescaled mode function follows as
\be
\widetilde\sigma_{k_l} (N) =  \frac{\widetilde\sigma_{k_l} (N_l)}{\[1-\frac{\lambda}{\nu_l^2}(N-N_1)\]^{1/4}} ~ {\rm exp}\(\int_{N_l}^{N} \sqrt{ \nu_l^2- \lambda (N'-N_l)} ~dN'\)~,
\ee
where
\ba \label{integral}
\int_{N_l}^{N} \sqrt{ \nu_l^2- \lambda (N' -N_l)} ~dN'
&=&  \frac{2}{3}\frac{1}{\lambda } \nu_l^3
- \frac{2}{3}\frac{1}{\lambda } \[\nu_l^2 - \lambda  (N-N_l) \]^{3/2} .
\ea
We can get the same solution by taking the asymptotic expansion of Airy functions in Eq.~\eqref{airy}.
The evolution of the rescaled mode function is shown in the left panel of Fig.~\ref{fig:running} for the negative running case, and in the left panel of Fig.~\ref{fig:running2} (the first 10 e-folds there) for $\lambda>0$. We see that, the WKB solutions agree with the full numerical results of Eq. \eqref{EoM}. Moreover, on superhorizon scales, they deviate from the results with constant masses, and thus are expected to modify the scaling behaviour in the squeezed bispectrum.

To show the phenomenological effects more explicitly, we further consider the situation with $ \nu_l^2 \gg |\lambda (N-N_l)|$, then the series expansion of \eqref{integral} yields
\ba
\nu_l \cdot (N-N_l) - \frac{1}{4\nu_l} \lambda    (N-N_l)^2 +...
\ea
Then the superhorizon decay of the isocurvature mode function is approximately given by
\be
\sigma_{k_l} (N) =  \sigma_{k_l} (N_l)  e^{-(3/2-\nu_l)(N - N_l) - \frac{1}{4\nu_l} \lambda  (N-N_l)^2 } ~.
\ee
Therefore one can easily get its amplitude at $N_s$ when the short wavelength modes exit the horizon.
Similar with the situation in \eqref{modulat}, the long mode modulation yields the squeezed limit accordingly, and here the scaling is modified to be
\begin{eBox}
\be \label{lightrun}
\lim_{k_l\ll k_s} B_\zeta \propto \frac{1}{k_l^3k_s^3} \(\frac{k_l}{k_s}\)^{3/2-\nu_l + \alpha_\nu \ln(k_s/k_l)} ~, ~~~~ {\rm with} ~
\alpha_\nu \equiv  \frac{\lambda}{4\nu_l} = \frac{1}{4\nu_l}   \epsilon_l\mpl^2\mathbb{R} \cdot  \eta_l .
\ee
\end{eBox}
We notice that the running index $\alpha_\nu$ leads the bispectrum to interpolate between the scalings given by the mass  $\mu_l$ and the mass $\mu(N_s)$.
When the curvature is positive, $\mu^2$ increases and $\alpha_\nu>0$.
If we fix $k_l$, then for small $k_s$ the bispectrum is closer to the local shape,  and it moves towards the equilateral scaling when $k_s$ increases.
For negative $\mathbb{R}$, the running of the scaling index with $k_s$ would be the opposite.

\subsection{Running in the $\mu>3H/2$ regime}

\begin{figure}[tbhp]
\centering
\includegraphics[width=0.46\linewidth]{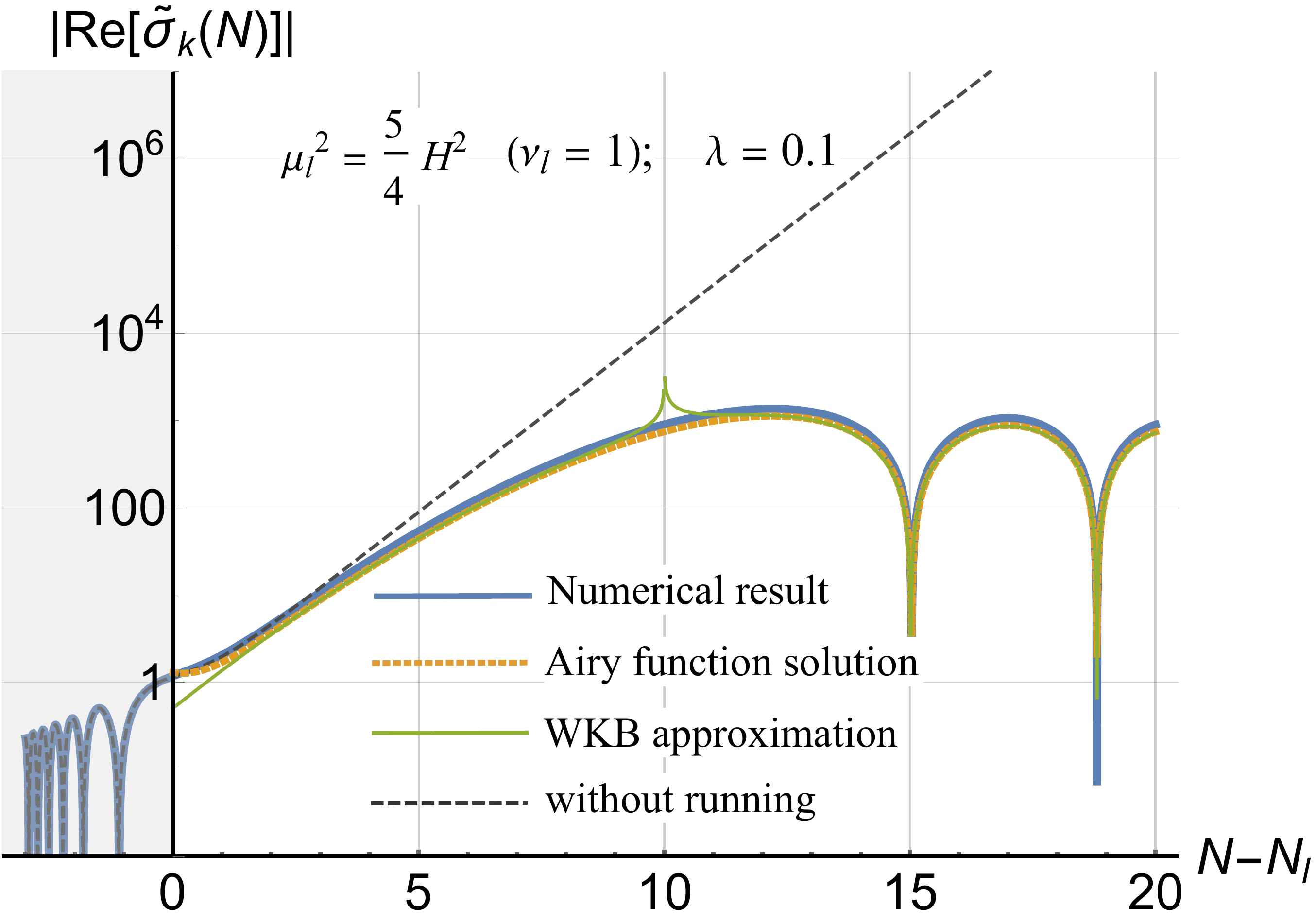}
\includegraphics[width=0.48\linewidth]{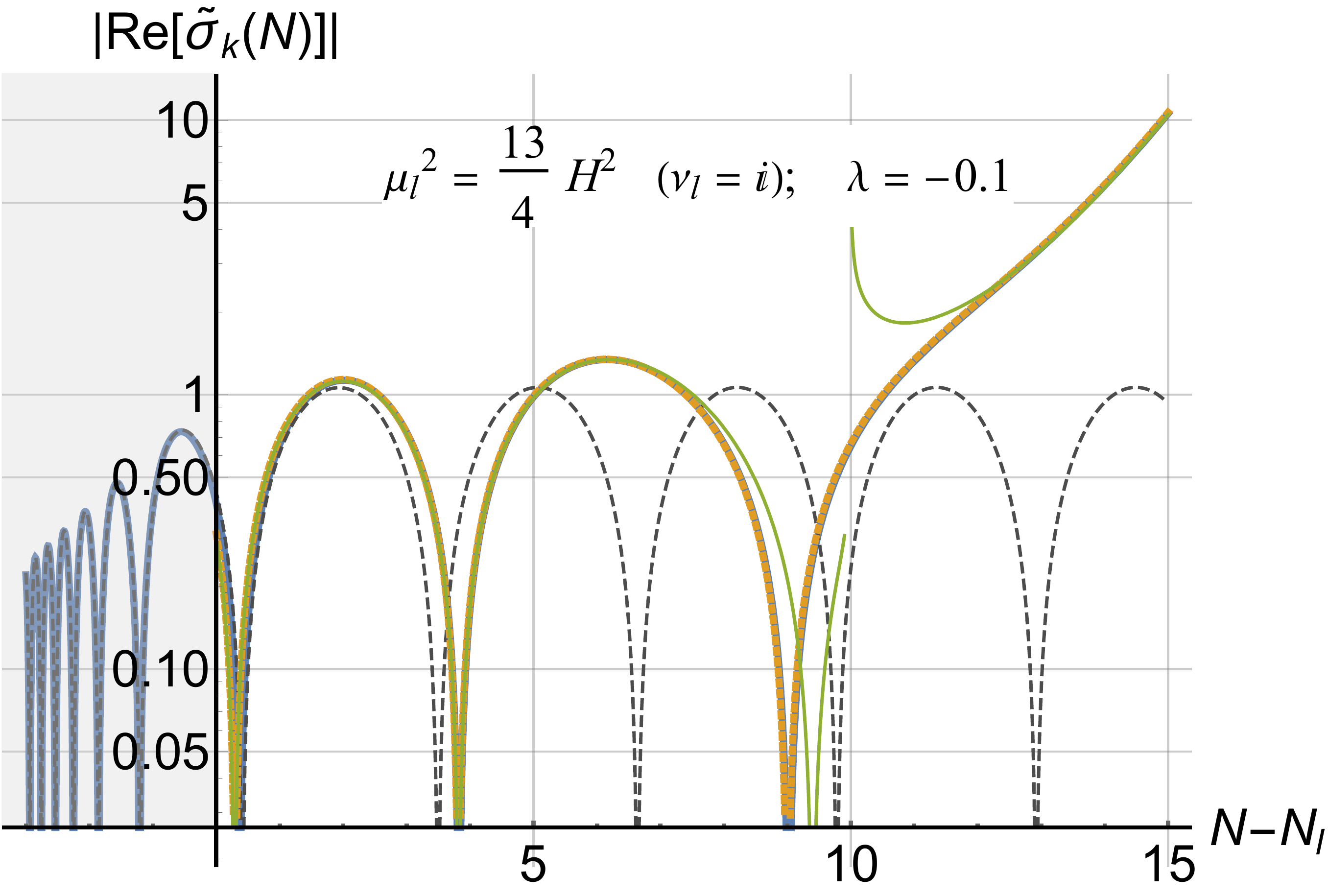}
\caption{The evolution of the rescaled isocurvature mode function for cases with: positive-running mass from $\mu<3H/2$ to $\mu>3H/2$ (left panel) and negative-running mass from $\mu>3H/2$ to $\mu<3H/2$ (right panel). In both figures, the blue curves are the full numerical solutions of $\tilde{\sigma}_k$, the orange dotted lines are the superhorizon solutions using Airy functions, the WKB approximations are given by the green curves, and the dashed grey lines describe the evolution with constant masses ($\lambda=0$). The shaded parts correspond to the subhorizon regime.}
\label{fig:running2}
\end{figure}

Next we turn to study the heavy field case. In this regime, $\nu_l$ is imaginary and the running isocurvature mass satisfies $\nu_l^2 - \lambda  (N-N_l)<0$.
Similarly we take the WKB approximation  and leave its violations for the next subsection, while here the adiabatic condition becomes
\be
|\nu_l^2 - \lambda  (N-N_l)| \gg (|-\lambda |)^{2/3}~.
\ee
Under this, \eqref{EoMr} yields the following oscillating WKB solutions
\be
\widetilde\sigma_{k_l} (N) \rightarrow  \frac{C_{\pm}}{\left|\nu_l^2-\lambda(N-N_1)\right|^{1/4}} ~ {\rm exp}\(\pm i \int_{N_l}^{N} \sqrt{ |\nu_l^2- \lambda (N-N_l)|} ~dN\)~,
\ee
whose real part can be further simplified into
\be
{\rm Re}~\widetilde\sigma_{k_l} (N) \propto
\frac{1}{\left|\nu_l^2-\lambda(N-N_1)\right|^{1/4}} ~
\cos\[ i\frac{2}{3\lambda } \left|\nu_l^2 - \lambda  (N-N_l) \right|^{3/2} -i\frac{2\nu_l^3}{3\lambda }  +\delta_l\]
~.
\ee
Here $\delta_l$ depends on $\nu_l$ and the initial conditions.
Again this can also be obtained by taking the asymptotic expansion of  Eq.~\eqref{airy}.
The result of the positive running mass is in the right panel of Fig.~\ref{fig:running}, and the right panel of Fig.~\ref{fig:running2} (the first 7 e-folds there) shows the ones for the negative running mass.
We find good agreement with numerical results.
As we can see, the positive running decreases the oscillation period while the negative running increases it.
This can be shown more clearly if we take $|\nu_l|^2 \gg |\lambda (N-N_l)|$ and expand the above solution. The superhorizon isocurvature mode function follows as
\be
{\rm Re}~\sigma_{k_l} (N) \propto e^{-\frac{3}{2}(N-N_l)}
\cos\[ i \nu_l (N - N_l) \(1 - \frac{\lambda}{4\nu_l^2} (N-N_l)\)   +\delta_l\]
~.
\ee
Again considering its modulation on the $k_s$-mode at $N_s$, we get the following  scaling behaviour in the squeezed bispectrum
\begin{eBox}
\be \label{heavyrun}
\lim_{k_l\ll k_s} B_\zeta \propto \frac{1}{k_l^3k_s^3} \(\frac{k_l}{k_s}\)^{3/2}\cos\[ i{\nu_l}\ln\(\frac{k_l}{k_s}\) -i \alpha_\nu  \ln^2\(\frac{k_l}{k_s}\) + \delta_l \]~,
~~ {\rm with} ~ \alpha_\nu \equiv  \frac{\lambda}{4\nu_l}~.
\ee
\end{eBox}
Notice here like $\nu_l$, the running index $\alpha_\nu$ is also imaginary and can be expressed as $\alpha_\nu = -i\epsilon_l\mpl^2\mathbb{R} \cdot  \eta_l /(4|\nu_l|)$.
Therefore due to the field space curvature, the oscillatory signal in the heavy field regime of QSFI would also be modified.

\subsection{Running through $\mu=3H/2$}
\label{sec:through}

Now we consider the situation where the WKB approximation breaks down. This corresponds to the cases when the isocurvature mass runs through  $\mu=3H/2$, and thus $\nu_l^2 - \lambda  (N-N_l) \simeq 0$.

Let us first take a look at the numerical results in Fig.~\ref{fig:running2}. For a positive $\lambda$, the isocurvature field runs from the light field regime to the heavy field regime, and the superhorizon behaviour of $\sigma_k$ demonstrates a smooth transition from the overdamped decay to the underdamped oscillation (left panel).
On the other hand, for the negative running, $\mu^2$ drops below the critical mass, and then the mode function transits from the oscillatory form to the exponential decay (right panel).

We can clearly see that the WKB solution becomes invalid when the mass runs close to $3H/2$. However, the analytical solutions \eqref{airy} with two Airy functions still holds true in this transition regime, and provides a good description for the mode function.
It is also interesting to notice that, the mathematics describing the transition between underdamped decay and overdamped oscillation is the same with the semi-classical approximation in quantum mechanics \cite{Sakurai:2011zz}, where the wave function is oscillating in the classically allowed region and decaying in the tunnelling regime. Therefore the critical mass $\mu=3H/2$ here can be seen as a ``turning point'' where the WKB approximation cannot be valid.

For the squeezed limit of the bispectrum, this transition behaviour of the superhorizon mode function may leave distinct imprints, with a combination of power-law and oscillatory signals.
Thus in this case, the deviation from the standard QSFI predictions could be large, and can be seen as a new template for the squeezed bispectrum. But for detectability, we need to be lucky such that the transition behaviour just occurs for the perturbation modes that correspond to our observational window. Or it is also possible that, future observations for different scales may help us to find the hint of this signature. For instance, if a oscillatory signal is detected by large scale experiments (such as CMB), while we observe power-law scaling of the squeezed bispectrum on small scales (such as LSS and CMB distortions), then it would indicate a negative running isocurvature mass caused by a hyperbolic-type field space.

\

In summary, the running of the isocurvature mass leads to the running in the scaling of the squeezed bispectrum, and the running index $\alpha_\nu$ measures the curvature of the field space.
With the above analysis, we close this section by giving two final remarks:

\begin{figure}[tbhp]
\centering
\includegraphics[width=0.8\linewidth]{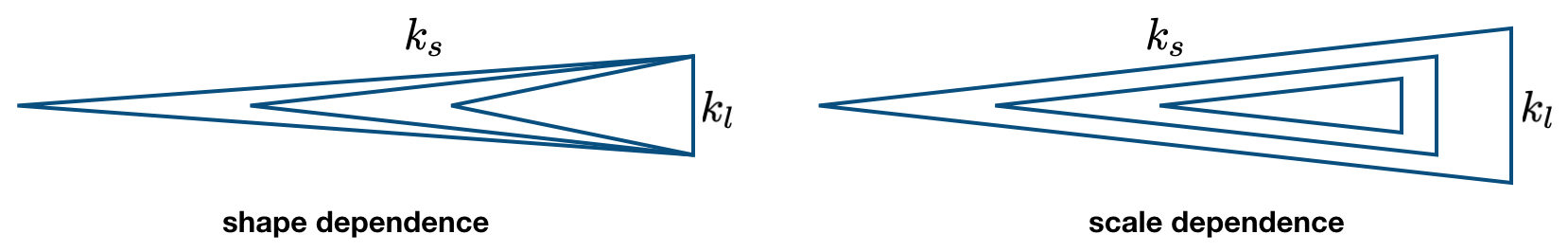}
\caption{Two types of squeezed configurations of momentum triangles. The bispectrum may only depend on the shapes (left), or there is dependence on the size as well (right).}
\label{fig:triangle}
\end{figure}

{\it Shape-dependence \& scale-dependence.}
In the standard QSFI with constant mass, one interesting fact is that, although the squeezed limit depends on the ratio of $k_l/k_s$, the full bispectrum is still scale-invariant.
That is to say, if we rescale three momenta but keep their ratio fixed, which corresponds to the transformation to a similar triangle, then the bispectrum remains unchanged.
Meanwhile the nontrivial scaling behaviour in \eqref{underscaling} and  \eqref{overscaling} can be seen as a function of the various squeezed configurations. For instance, if we fix the long wavelength mode $k_l$, and let $k_s$ vary, the bispcetrum becomes different. Thus this result is shape-dependent.
For a running isocurvature mass, as we can see from \eqref{lightrun} and \eqref{heavyrun} the shape dependence remains. Furthermore the bispectrum here also becomes scale-dependent.
If we rescale three wavenumbers together, for instance $k_l \rightarrow \kappa k_l$ and $k_s \rightarrow \kappa k_s$, then the new scaling index $\nu_\kappa$ should be determined by the $\mu_\kappa^2$ when the rescaled long wavelength mode $\kappa k_l$ exits the horizon
\be
\nu_\kappa= \sqrt{\frac{9}{4}-\frac{\mu_\kappa^2}{H^2}} \simeq \nu_l - \frac{\lambda}{2\nu_l} (N_\kappa - N_l) ~.
\ee
Notice that the scale-dependent running index here $\tilde\alpha_\nu \equiv d \nu_l / (d N)=-\lambda / (2\nu_l)$ differs from the ones parametrizing shape-dependence in \eqref{lightrun} and \eqref{heavyrun}.
This is because, a running isocurvature mass would continuously affect the superhorizon evolution of $\sigma_{k_l}$, which plays an important role for the shape-dependence; while the scale-dependence is controlled by the mass at different times, thus the running can be simply obtained by taking derivative of $\mu^2$.

{\it Implications on the scale-dependent bias.} In the LSS observations,  the halo overdensity $\delta_h$ tracing galaxy distribution and the dark matter density contrast $\delta_m$ are related by the bias $b$ through $\delta_h=b\delta_m$.
It has been shown that, signals in the squeezed limit of the primordial scalar bispectrum can lead to a scale-dependent component in the bias $b^{\rm NG}$, which becomes dominant on large scales \cite{Dalal:2007cu, Matarrese:2008nc} (see Ref. \cite{Biagetti:2019bnp} for a recent review).
Thus this scale-dependent halo bias provides an observational opportunity for the detection of the modified scaling behaviour studied above.
In the conventional QSFI models, the  scaling signals in \eqref{underscaling} and  \eqref{overscaling} respectively imply $b^{\rm NG} (k) \propto k^{-1/2-\nu}$
and  $b^{\rm NG} (k) \propto k^{-1/2}\cos (i\nu\ln k)$. Accordingly the squeezed bispectra with running in \eqref{lightrun} and \eqref{heavyrun} yield
\be
b^{\rm NG} (k) \propto {k^{-1/2- \nu_l - \alpha_\nu \ln k}}
~~~~ {\rm and} ~~~~  b^{\rm NG} (k) \propto {k^{-1/2}}\cos (i\nu_l\ln k - i\alpha_\nu \ln^2 k)~,
\ee
which suggests that, besides $\nu$, the running index $\alpha_\nu$ can also be set as a free parameter for the data analysis of future LSS surveys.

\section{Conclusion and discussion}
\label{sec:concl}

In this paper we explore the implications of nontrivial internal spaces in the context of inflationary massive fields. 
Here QSFI is generalized to curved field manifold, and then analyzed by using both the multi-field techniques and the background EFT approach.
Through the multi-field analysis, we show that the field space curvature could contribute significantly to the isocurvature mass in QSFI, thus modify its predictions on non-Gaussianity.
Meanwhile the same result is also derived in the EFT of the background fields, where  a dim-6 operator is identified to generate the same effects as the curved field space.
We build the connection between these two different but equivalent approaches, and further demonstrate that the cutoff scale of the dim-6 operator is associated with the curvature scale of the field space.

Moreover, as a result of the slow-roll dynamics of the inflaton field, the field space curvature contribution to the isocurvature mass is time-dependent in nature.
We perform the first analysis on phenomenological consequences of the running isocurvature mass, and find new features in the scaling of the squeezed scalar bispectrum.
Besides the power-law and oscillatory signals of QSFI in the light and heavy mass regimes, the time-dependence of the isocurvature mass leads to running behaviour in the squeezed scaling.
{If the field space is positively curved, the isocurvature mass increases, which leads to the positive running in the squeezed scaling.
While for the field space with negative curvature, the running becomes negative.}
Also a transition signal between the power-law and oscillatory scalings is discovered when the mass runs through $\mu=3H/2$.
{These modifications to the previous results of QSFI provide new templates for detecting primordial non-Gaussianity.
Therefore in future observations, through the precise measurement of running behaviours in the squeezed bispectrum, we may be able to probe the geometry of the internal field space during inflation.}

This work can also be seen as the first step towards several possible directions for future research.
First of all, it is interesting to study the implications on cosmological collider physics \cite{Arkani-Hamed:2015bza}.
Our results indicate that, due to the time-dependent background of the inflaton field, higher dimension operators may become non-negligible for the collider signals. While the current work can be directly applied to heavy scalar particles, it is worth investigating similar effects of particles with spins, whose ``bare'' mass may also be corrected by higher order EFT operators mixing with the inflaton.

Next, considering that one of our main motivations is to probe field space curvature during inflation model-independently, the current results are not sufficiently general and unique yet.
For instance, in principle it is possible to engineer other models of QSFI with running isocurvature mass, which would lead to similar phenomenology degenerate with the geometrical effects\footnote{The analysis of the time-dependent isocurvature mass starting from the parametrization \eqref{runningmass} can be seen as an independent part of this paper, which is also of phenomenological interest in contexts beyond  curved field space.
Though one may wonder if other constructions are as simple and natural as the one considered here.}.
Thus we are encouraged to explore the truly unique  signatures of the curved field space with more generalities.

Finally, the running scaling signals in the squeezed limit of the scalar bispectrum have implications for observations, which deserve a closer look.
For example, the observability of these signals and the fitting of the running index using CMB and LSS data remain to be investigated.

\acknowledgments

I would like to thank Ana Ach\'ucarro, Guilherme Pimentel and Yvette Welling for collaborating on related projects and comments on the draft.
I am also grateful to
Yi-Fu Cai, Xingang Chen,
S\'ebastien Renaux-Petel
and Yi Wang for valuable discussions and comments.
This work has benefited from stimulating discussions during the ``Inflation and Geometry'' Workshop at IAP Paris and the Beyond Conference in Warsaw.
DGW is supported by a de Sitter Fellowship of the Netherlands Organization for Scientific Research (NWO).


\bibliographystyle{JHEP}%
\bibliography{bibfile}

\end{document}